\documentclass[fleqn,usenatbib]{mnras}

\usepackage{newtxtext,newtxmath}

\usepackage[T1]{fontenc}
\usepackage{ae,aecompl}

\usepackage{graphicx}	
\usepackage{amsmath}	
\usepackage{amssymb}	

\usepackage{hyperref} 
\hypersetup{
    colorlinks,
    citecolor=blue, 
    filecolor=black,
    linkcolor=blue, 
    urlcolor=red
}
\usepackage{geometry}
\usepackage{pdflscape}
\usepackage{subfig}
\usepackage{longtable}

\usepackage{tablefootnote}

\usepackage{threeparttable}

\title[Radio galaxy at z = 5.72]{Discovery of a radio galaxy at z = 5.72}

\author[A. Saxena et al.]{A. Saxena$^{1}$\thanks{E-mail: saxena@strw.leidenuniv.nl}, M. Marinello$^{1, 2}$, R. A. Overzier$^{2}$, P. N. Best$^{3}$, H. J. A. R{\"o}ttgering$^{1}$,
\newauthor K. J. Duncan$^{1}$, I. Prandoni$^{4}$, L. Pentericci$^{5}$, M. Magliocchetti$^{6}$, D. Paris$^{5}$, F. Cusano$^{7}$,
\newauthor F. Marchi$^5$, H. T. Intema$^{1}$ and G.K. Miley$^{1}$ \\ \\
$^{1}$Leiden Observatory, Leiden University, P.O. Box 9513, 2300 RA Leiden, The Netherlands\\
$^{2}$Observat\'{o}rio Nacional, Rua General Jos\'{e} Cristino, 77, S\~{a}o Crist\'{o}v\~{a}o, Rio de Janeiro, RJ, CEP 20921-400, Brazil\\
$^{3}$Institute for Astronomy, University of Edinburgh, Royal Observatory, Blackford Hill, EH9 3HJ Edinburgh, UK\\
$^{4}$INAF-Instituto di Radioastronomia, Via P. Gobetti 101, I-40129 Bologna, Italy\\
$^{5}$INAF-Osservatorio Astronomico di Roma, Via Frascati 33, I-00040 Monteporzio (RM), Italy\\
$^{6}$IAPS-INAF, Via Fosso del Cavaliere 100, I-00133 Rome, Italy\\
$^{7}$INAF-Osservatorio di Astrofisica e Scienza dello Spazio di Bologna, Via P. Gobetti 93/3, I-40129 Bologna, Italy
}

\date{Accepted 2018 July 12. Received 2018 July 5; in original form 2018 June 4}

\pubyear{2018}

\begin{document}
\label{firstpage}
\pagerange{\pageref{firstpage}--\pageref{lastpage}}
\maketitle

\begin{abstract}
We report the discovery of the most distant radio galaxy to date, TGSS J1530+1049 at a redshift of $z=5.72$, close to the presumed end of the Epoch of Reionisation. The radio galaxy was selected from the TGSS ADR1 survey at 150 MHz for having an ultra-steep spectral index, $\alpha^{\textrm{\tiny{150 MHz}}}_{\textrm{\tiny{1.4 GHz}}} = -1.4$ and a compact morphology obtained using VLA imaging at 1.4 GHz. No optical or infrared counterparts for the radio source were found in publicly available sky surveys. Follow-up optical spectroscopy at the radio position using GMOS on Gemini North revealed the presence of a single emission line. We identify this line as Lyman alpha at $z=5.72$, because of its asymmetric line profile, the absence of other optical/UV lines in the spectrum and a high equivalent width. With a Ly$\alpha$ luminosity of $5.7 \times 10^{42}$ erg s$^{-1}$ and a FWHM of $370$ km s$^{-1}$, TGSS J1530+1049 is comparable to `non-radio' Lyman alpha emitters (LAEs) at a similar redshift. However, with a radio luminosity of $\log L_{\textrm{150 MHz}} = 29.1$ W Hz$^{-1}$ and a deconvolved physical size $3.5$ kpc, its radio properties are similar to other known radio galaxies at $z>4$. Subsequent $J$ and $K$ band imaging using LUCI on the Large Binocular Telescope resulted in non-detection of the host galaxy down to $3\sigma$ limits of $J>24.4$ and $K>22.4$ (Vega). The $K$ band limit is consistent with $z>5$ from the $K-z$ relation for radio galaxies and helps rule out low redshifts. The stellar mass limit derived using simple stellar population models is $M_{\textrm{stars}}< 10^{10.5}$ $M_\odot$. Its relatively low stellar mass and small radio and Ly$\alpha$ sizes suggest that TGSS J1530+1049 may be a radio galaxy in an early phase of its evolution. 
\end{abstract}




\section{Introduction}

High-redshift radio galaxies (HzRGs) are thought to be the progenitors of local massive elliptical galaxies and generally contain large amounts of dust and gas \citep{bes98b, car02b, reu04, deb10}. They are also among the most massive galaxies at their redshift \citep{ove09} and are often found to be located at the centre of clusters and proto-clusters of galaxies \citep{pen00, ven02, rot03, mil04, hat11, ors16}. Studies of their environment can give insights into the assembly and evolution of the large scale structure in the Universe. \citet{mil08} provide an extensive review about the properties of distant radio galaxies and their environments.

Radio galaxies at $z>6$, in the Epoch of Reionisation (EoR), are of particular interest as they could be used as unique tools to study the process of reionisation in detail. At these redshifts, the hyper-fine transition line from neutral hydrogen atoms, with a rest-frame wavelength of 21 cm, falls in the low-frequency radio regime and can be observed as absorption signals in spectra of luminous background radio sources such as radio galaxies \citep{car02, fur02, xu09, mac12, ewa14, cia15}. Such 21 cm absorption signals from patches of neutral hydrogen clouds in the early Universe could in principle be observed by current and next-generation radio telescopes such as the Giant Metre-wave Radio Telescope (GMRT; \citealt{gmrt}), the Low Frequency Array (LOFAR; \citealt{lofar}), the Murchinson Widefield Array (MWA; \citealt{mwa}) and the Square Kilometer Array (SKA; \citealt{ska}). This unique application motivates searches for bright enough radio galaxies at the highest redshifts from deep all-sky surveys at low radio frequencies.

Finding powerful radio galaxies at increasingly large distances or redshifts, however, is challenging. HzRGs are rare and flux-limited samples have shown that their space densities fall off dramatically at $z>2-3$ \citep{dun90, wil01, rig11, rig15}. Although a number of quasars at $z>5$ are known, with a few also being radio-loud \citep[see][for example]{ban15}, the same cannot be said for radio galaxies - previously, the only known radio galaxy at $z>5$ was TN J0924$-$2201 at $z=5.19$ \citep{vbr99}. If the orientation-based unification of radio galaxies and quasars is valid \citep[see][for example]{mor17}, the number of radio quasars and galaxies at any given epoch should be comparable and it may be possible that many of the already unidentified radio sources are at $z>5$. The scarcity of $z>5$ radio galaxies could therefore be due to the relative difficulty in first identifying these sources amongst the wider radio source population, and then obtaining spectroscopic redshifts for these radio galaxies, which are optically much fainter than quasars. Dedicated spectroscopic follow-up of radio sources such as the WEAVE-LOFAR survey \citep{smi16}, and the upcoming major optical facilities such as the Extremely Large Telescope (ELT), the Thirty Meter Telescope (TMT), the Giant Magellan Telescope (GMT) and the James Webb Space Telescope (JWST) will help overcome these difficulties and potentially help characterise a number of USS radio sources.

The key requirement for gathering enough statistics for meaningful studies of radio galaxies at high redshifts are deep low-frequency radio surveys covering large areas on the sky. Surveys such as the TIFR GMRT Sky Survey Alternative Data Release 1 (TGSS; \citealt{int17}) and the currently ongoing surveys using LOFAR \citep{shi17} are opening up new parameter spaces for searches for radio galaxies at $z\ge6$. Using TGSS, which covers the entire radio sky north of $-53$ declination at a frequency of 150 MHz and achieving a median noise level of $3.5$ mJy beam$^{-1}$, we launched a campaign to hunt for fainter and potentially more distant HzRGs, with the ultimate aim of discovering radio galaxies that could be suitable probes of the EoR \citep{sax18}. In this paper, we report the discovery of a radio galaxy at a redshift of $z=5.72$, TGSS J1530+1049, which was pre-selected as part of our sample of high-redshift radio galaxy candidates.

The layout of this paper is as follows. In Section \ref{sec:data} we present details about the initial source selection criteria and  follow-up radio observations at high resolution for TGSS J1530+1049. In Section \ref{sec:obs} we present the new optical spectroscopy and infrared imaging obtained for TGSS J1530+1049 and expand upon the data reduction methods. In Section \ref{sec:redshift} we describe how the redshift for this source was determined. In Section \ref{sec:discussion} we study the emission line and radio properties of this source and set constraints on its stellar mass. We also compare the observed properties to galaxies at the same epoch from the literature. Finally, in Section \ref{sec:conclusion} we summarise the findings of this paper. Throughout this paper we assume a flat $\Lambda$CDM cosmology with $H_0$ = 70 km s$^{-1}$ Mpc$^{-1}$ and $\Omega_m = 0.3$. Using this cosmology, at a redshift of 5.72 the age of the Universe is 0.97 Gyr, and the angular scale per arcsecond is 5.86 kpc.

\section{Source selection}
\label{sec:data}
Our two stage selection process is based on first isolating compact radio sources with an ultra-steep spectrum (USS; $\alpha<-1.3$, where $S_\nu \propto \nu^\alpha$) at radio wavelengths, that has historically been very successful at finding HzRGs from wide area radio surveys \citep{rot94, blu99, deb00, afo11}, and then combining it with optical and/or infrared faintness requirements. The relation that exists between the apparent $K$-band magnitude of radio galaxies and their redshift, known as the $K-z$ relation, \citep{lil84, jar01, wil03, roc04} gives further strength to the argument of selecting USS sources that are also faint at near-infrared wavelengths in a bid to isolate HzRGs \citep{ker12}. Deep near-infrared imaging of promising USS candidates can therefore serve as an independent way to set constraints on the redshifts of radio sources. HzRGs are expected to be very young and therefore, have small sizes at the highest redshifts \citep{sax17}: implementing an additional criterion that puts an upper limit on the angular sizes of radio sources has the potential to increase the efficiency of pin pointing the highest-redshift sources in a wide area radio survey.

Combining all of these selection methods, we first shortlisted 588 candidates with an ultra-steep spectrum ($\alpha^{\textrm{\tiny{150 MHz}}}_{\textrm{\tiny{1.4 GHz}}}<-1.3$) and compact morphologies, out of a total of 65,996 sources that had spectral index information from TGSS and FIRST/NVSS. This sample probes fainter flux densities than previous large area searches and has flux limits that ensure that a new parameter space in flux density and spectral index is probed, where a large number of undiscovered HzRGs are expected to lie \citep{ish10}. From this shortlist, we then retained in our sample only those radio sources that are blank in all available optical surveys such as the Sloan Digital Sky Survey DR12 (SDSS; \citealt{ala15}) and the Pan-STARRS1 survey (PS1; \citealt{cha16}), and infrared surveys such as ALLWISE using the \textit{WISE} satellite \citep{wri10} and the UKIDSS surveys \citep{law07} to maximise the chances of finding radio galaxies at the highest redshifts. This led to a final sample of 32 very promising HzRG candidates. Details of the sample selection can be found in \citet{sax18}.
\begin{figure}
\centering
	\vspace{-10pt}
	\includegraphics[scale=0.45]{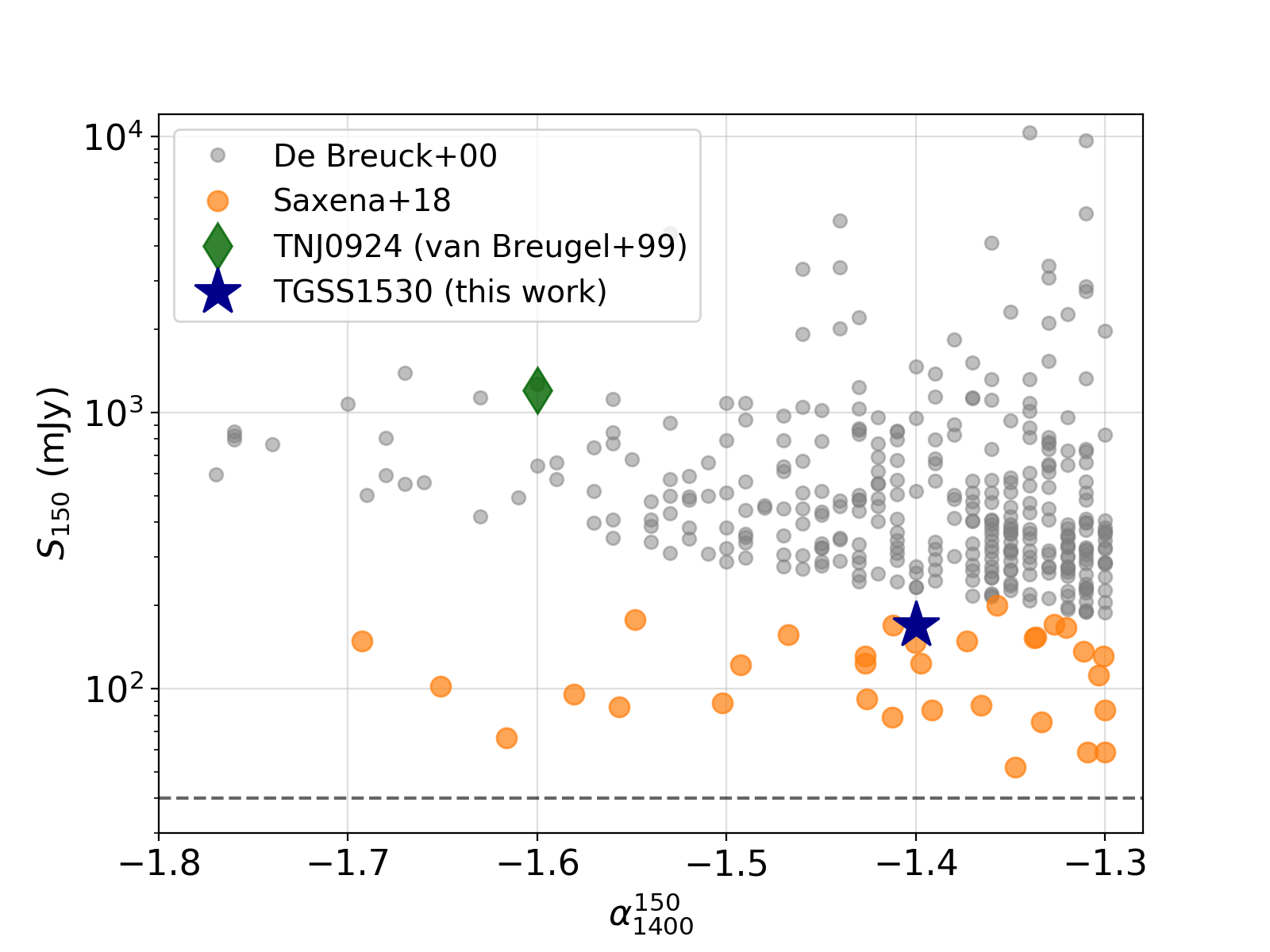}
    \caption{The location of TGSS J1530+1049 in the flux density$-$spectral index parameter space. The large orange points show the parameter space probed by the \citet{sax18} sample and the smaller grey points show radio sources from \citet{deb00}, scaled to an observed frequency of 150 MHz using the spectral indices provided for individual sources. Also shown for comparison is TN J0924$-$2201 at $z=5.2$ \citep{vbr99}. TGSS J1530+1049 is fainter than the previously studied large area samples and offers a new window into fainter radio galaxies at high redshifts.}
    \label{fig:fluxalpha}
\end{figure}

High resolution imaging using the Karl G. Jansky Very Large Array (VLA) for the 32 candidates, including for TGSS J1530+1049 (RA: 15:30:49.9, Dec: $+$10:49:31.1) is presented in \citet{sax18}, which was used to obtain morphologies and sub-arcsecond localisation of the expected positions of the host galaxies, enabling blind spectroscopic follow-up. TGSS J1530+1049 in particular showed a compact morphology, which was fitted with a single Gaussian. With a flux density of $S_{\textrm{\tiny{150 MHz}}} = 170 \pm 34$ mJy, TGSS J1530+1049 is one of the brightest sources in the sample. At 1.4 GHz, it has a flux density $S_{\textrm{1.4 GHz}} = 7.5 \pm 0.1$ mJy, giving a spectral index of $\alpha = -1.4 \pm 0.1$. With a relatively small (deconvolved) angular size of $0.6 \pm 0.1$ arcsec, TGSS J1530+1049 was deemed to be a promising HzRG candidate. We show the location of TGSS J1530+1049 in the flux density$-$spectral index parameter space in Figure \ref{fig:fluxalpha}. 

\begin{figure}
\centering
\vspace{-10pt}
	\includegraphics[scale=0.16]{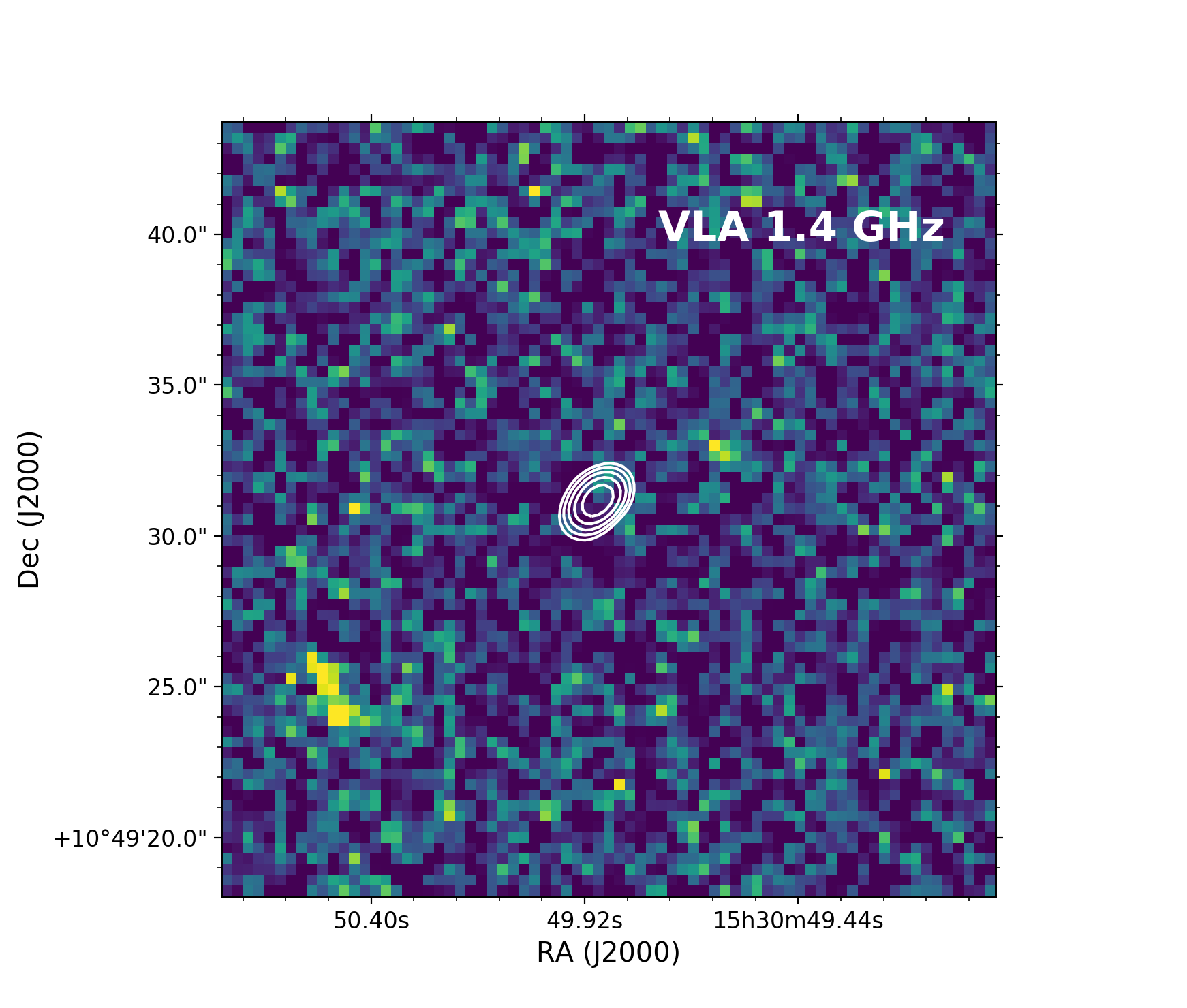}
	\vspace{-20pt}
	\caption{Stacked $y$, $J$, $H$ and $K$ band image from the UKIDSS Large Area Survey, with contours (starting from 0.5 mJy, in a geometric progression of $\sqrt{2}$) from the 1.4 GHz VLA map \citep{sax18} overplotted for TGSS J1530+1049. The radio source is compact and has an ultra-steep spectral index. A non-detection in the UKIDSS LAS $K$ band down to a magnitude limit of 18.4 Vega ($\sim 20.3$ AB) made TGSS J1530+1049 a promising HzRG candidate and a prime target for spectroscopic follow-up.}
	\label{fig:LAS}
\end{figure}

TGSS J1530+1049 is not detected in any of the PS1 bands ($g$, $r$, $i$, $z$ and $y$). This source also happens to lie in the sky area covered by the UKIDSS Large Area Survey (LAS), and is not detected down to (Vega) magnitude limits of $y>20.5$, $J>20.0$, $H>18.8$ and $K>18.4$. We show the image obtained from stacking all of the LAS bands with radio contours overlaid in Figure \ref{fig:LAS}. Lastly, this source is also not detected in any of the ALLWISE bands. These non-detections coupled with the ultra-steep radio spectral index and compact radio morphology are in line with expectations of a high-redshift host galaxy and made TGSS J1530+1049 a prime candidate for follow-up spectroscopy.

\section{Observations}
\label{sec:obs}
\subsection{Gemini GMOS spectroscopy}
A long-slit spectrum of TGSS J1530+1049TGSS J1530+1049 was taken using GMOS on Gemini North on 28 April, 2017 (Program ID: GN-2017A-Q-8; PI: Overzier) using the filter GG455\_G0305 and the R400\_G5305 grating giving a wavelength coverage of $5400-10800$ \AA$ $ and a resolution of roughly $R\sim1500$. The central wavelength was set to 7000 \AA$ $. The total length of the slit was 300 arcseconds and the slit width was chosen to be 1.5 arcseconds so that it covers the entire radio emission footprint detected in the VLA image. As the host galaxy of the radio source was undetected in all available all-sky optical/IR surveys, we performed blind offsetting from a bright star, which ensures positional accuracy to within 0.1 arcseconds, to the centroid of the radio emission. The VLA observations ensured sub-arcsecond localisation of the expected position of the host galaxy and the relatively large slit-width provided insurance against minor positional uncertainties. We took 3 exposures of 800 seconds each, giving a total of 2400 seconds of on-source exposure time. The standard star EG131 was observed for flux calibration.

We used the Gemini IRAF package for reducing the data, which includes the standard steps for optical spectrum reduction. Briefly, the bias frames were mean stacked in a master bias which was subtracted from all other images acquired. Pixel-to-pixel sensitivity was corrected through the flat field image taken during the day of the observations. The wavelength solution was derived from the arc lamp frame taken immediately after the science observations, and applied to the science frame and standard star. The 2D images were then combined in a single frame, rejecting possible cosmic rays. The sky lines were removed and flux calibration was achieved using the standard star spectrum.

\begin{figure*}
\centering
	\includegraphics[scale=0.25]{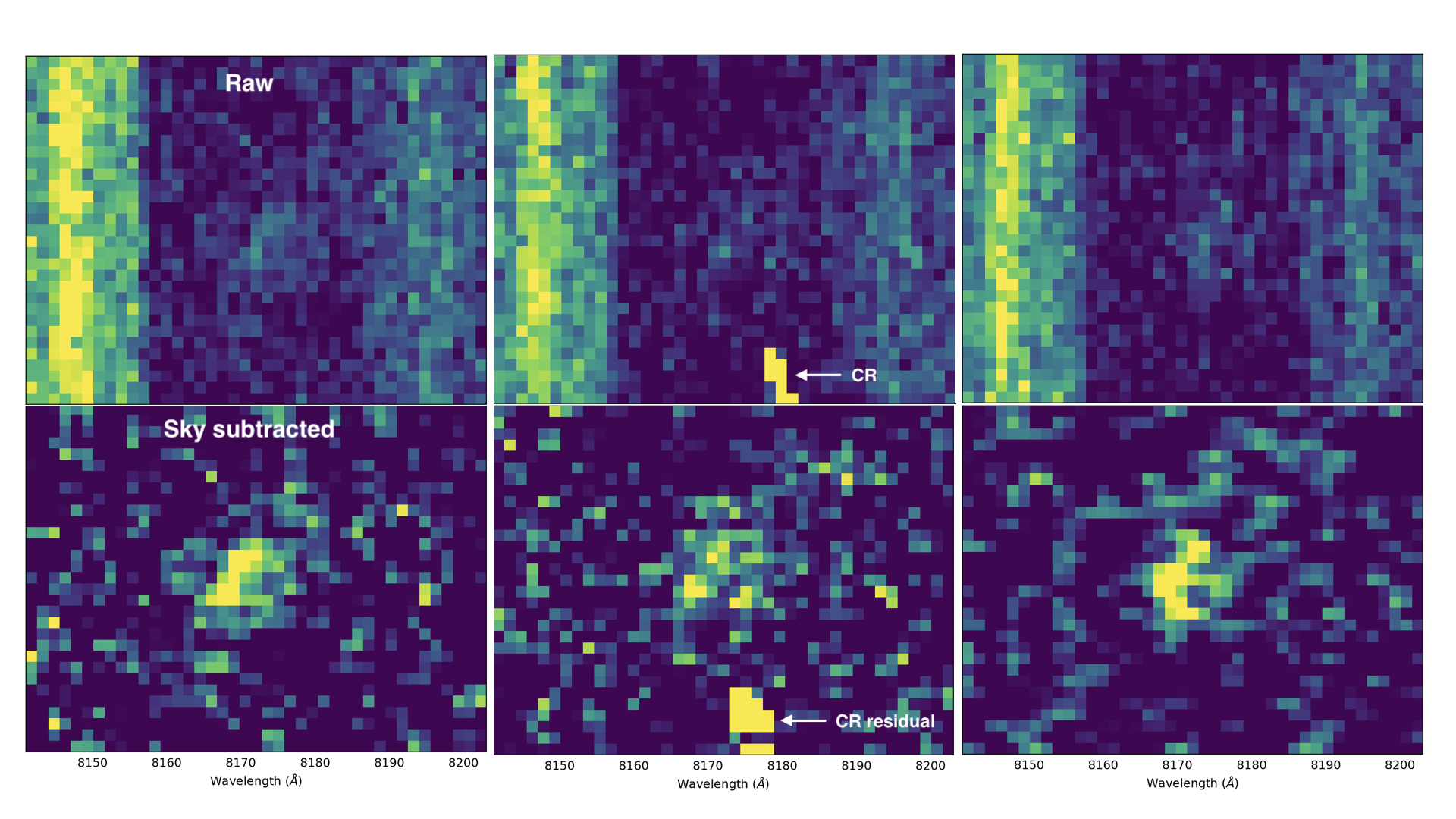}
    \caption{Raw (top panels) and sky subtracted (bottom panels) 2D frames shown for the three individual exposures taken using GMOS on Gemini. Traces of the emission line are visible in all three frames, ensuring that the detected line is real and not a consequence of cosmic rays or artefacts. There is some cosmic ray residual left over in the second frame but that does not contaminate the emission line signal.}
    \label{fig:allframes}
\end{figure*}

\begin{figure}
\centering
	\includegraphics[scale=0.5]{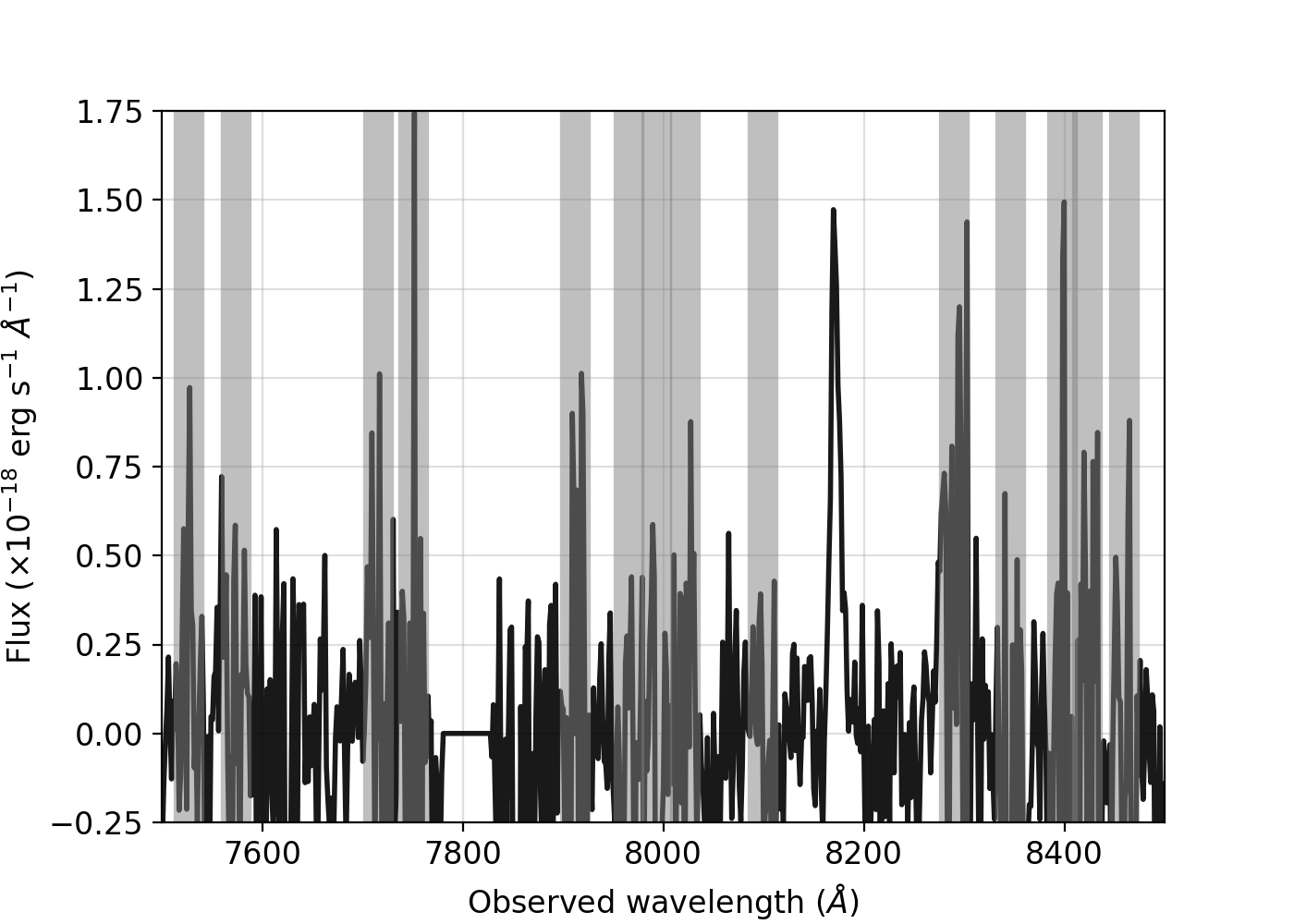}
	\vspace{-10pt}
	\caption{Extracted 1D spectrum showing the single emission line detection centred at 8170 \AA$ $ from the GMOS 2D spectrum. No other line or continuum is detected. Shaded regions mark the presence of sky lines in the spectrum.}
	\label{fig:pub_spec}
\end{figure}

A single emission line with a peak at 8170 \AA$ $ and a spatial extent of $\sim1$ arcsecond was detected in the reduced 2D spectrum at the expected position of the radio galaxy. No other line associated with this source was detected. No continuum was detected either bluewards or redwards of this line either. To ensure that the line detection is indeed real and not due to an artefact or contamination by cosmic rays, we looked at the individual frames, both raw and sky subtracted, to ensure that the detection (although marginal) was present in each science frame. The three frames are shown in Figure \ref{fig:allframes}. The top panels show the raw frames and the bottom panels the sky subtracted frames. The emission line is clearly present in all three frames, ensuring that the detection is real. The extracted 1D spectrum with a 1 arcsecond aperture showing the detected emission line is shown in Figure \ref{fig:pub_spec}. We give details about line identification in Section \ref{sec:redshift}.

\subsection{Large Binocular Telescope NIR imaging}
Imaging in the $J$ and $Ks$ bands using LUCI (formerly known as LUCIFER; \citealt{sei03}) on the Large Binocular Telescope (LBT) was carried out in two separate runs, with the first on 1 February 2018 and the second on 11 May 2018 (Program ID 2017\_2018\_43; PI: Prandoni). The average seeing throughout the observations was $0.6-0.8$ arcseconds. In the first run, the on-source exposure time was 720 ($12\times60$s) seconds in $J$ (central wavelength of 1.247 microns) and 1200 ($20\times60$s) seconds in $Ks$ (central wavelength of 2.194 microns). In the second run we obtained additional 3600 ($30\times120$s) seconds in $J$ and 3000 ($50\times60$s) seconds in $Ks$, giving a total on-source exposure time of 4320 seconds in $J$ band and 4200 seconds in $Ks$ band. 

The LUCI data reduction pipeline developed at INAF-OAR was used to perform the basic reduction such as dark subtraction, bad pixel masking, cosmic ray removal, flat fielding and sky subtraction. Astrometric solutions for individual frames were obtained and the single frames were then resampled and combined using a weighted co-addition to form a deeper image. The $4' \times 4'$ field-of-view of LUCI contained many bright objects detected in both 2MASS and the UKIDSS Large Area Survey, which were used to calibrate the photometry of the images in both bands.

\begin{figure}
\centering
\vspace{-10pt}
	\includegraphics[scale=0.19]{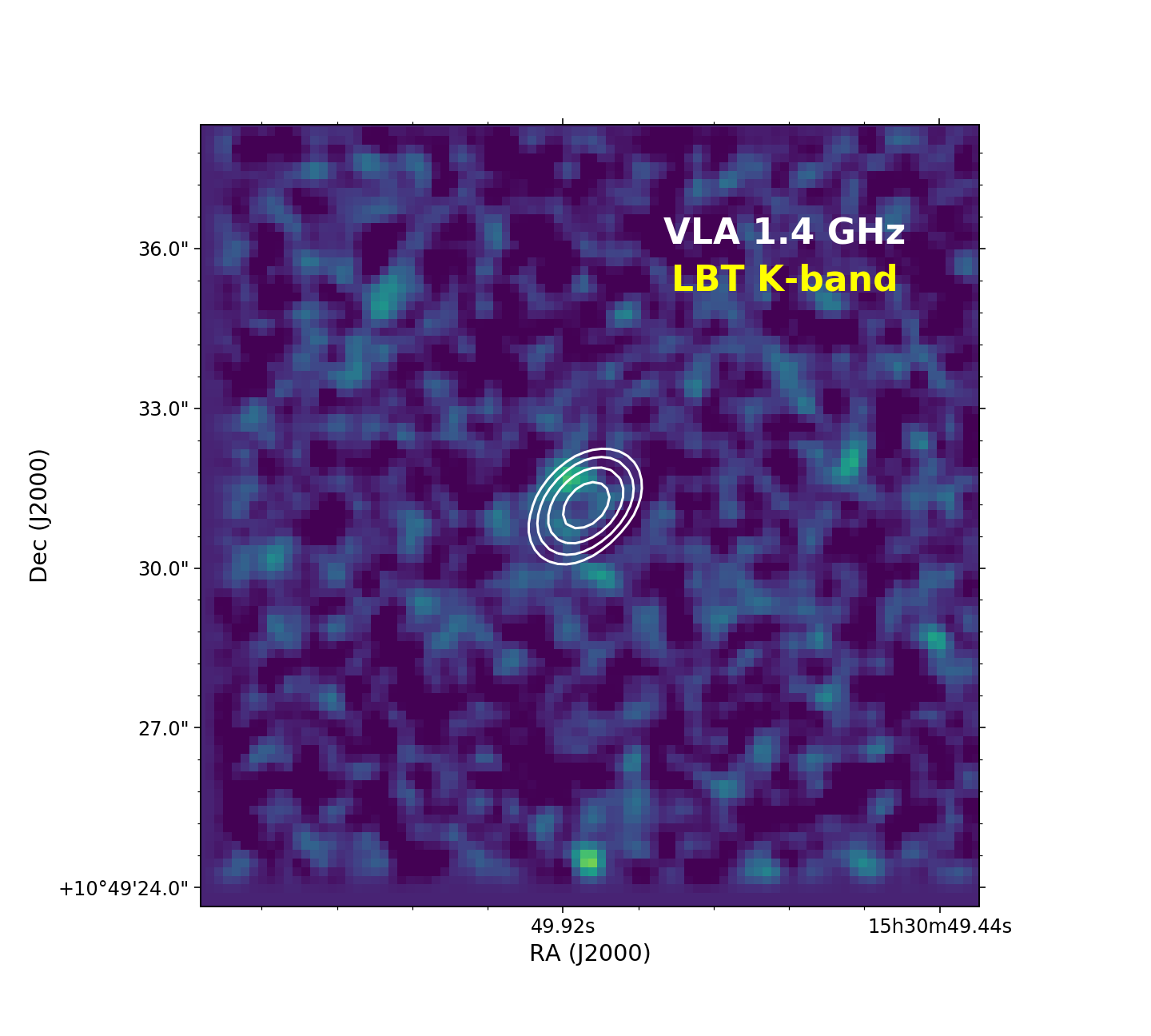}
	\vspace{-20pt}
	\caption{$K$-band image from the Large Binocular Telescope (LBT), which has been smoothed with a $3\times3$ pixel Gaussian kernel, with radio contours (same as Figure 2) from the VLA at 1.4 GHz overlaid. The measured magnitude at the radio position with a 1.5 arcsecond aperture is fainter than the $3\sigma$ depth of the image, giving $K>22.4$. When the image is smoothed, however, faint emission is visible around the peak of the radio emission. The magnitude limit is consistent with $z>5$ following the $K-z$ relation for radio galaxies. For comparison, the $z=5.2$ radio galaxy TN J09224$-$2201 has a K-band magnitude of 21.3 \citep{vbr99}.}
	\label{fig:kband}
\end{figure}

The median and standard deviation of the background in both images was calculated by placing 5000 random apertures with a  diameter of 1.5 arcseconds. We measure $3\sigma$ depths of $J = 24.4$ and $Ks=22.4$. Aperture photometry performed on both $J$ and $Ks$ (from here on we denote $Ks$ as simply $K$) images using \textsc{photutils} \citep{photutils} at the peak of the radio emission using an aperture of diameter 1.5 arcseconds yield magnitudes that are lower than the $3\sigma$ depths in both images. Smoothing the $K$ band image with a $3\times3$ pixel Gaussian kernel reveals a faint source very close to the peak radio pixel, as shown in Figure \ref{fig:kband}, but it is not entirely clear if this indeed the host galaxy and there is no faint detection even in the smoothed $J$ band image. A summary of the observations is given in Table \ref{tab:obs}.
\begin{table}
\centering
\caption{Observation log.}
	\begin{tabular}{l c c r}
	\hline
	Telescope	&		Instrument		&	Date			&	Exp. time (sec)	\\
	\hline \hline
	Gemini N	&	GMOS long-slit		&	28-04-2017	&	2400 ($3\times800$s)		\\
    \hline
	LBT			&	LUCI J-band		&	01-02-2018	&	720 ($12\times60$s)		\\
    			&					&	09-05-2018 	&	3600 ($30\times120$s)	\\ 
                &	Total			&		&	4320	\\
    \hline
	LBT			&	LUCI Ks-band		&	01-02-2018	&	1200 ($20\times60$s)	\\
    			&					&	09-05-2018	&	3000 ($50\times60$s)	\\
                &	Total			&				&	4200 \\
	\hline
	\end{tabular}
\label{tab:obs}
\end{table}	

\section{Redshift determination}
\label{sec:redshift}
\begin{figure}
\centering
	\vspace{10pt}
	\includegraphics[scale=0.215]{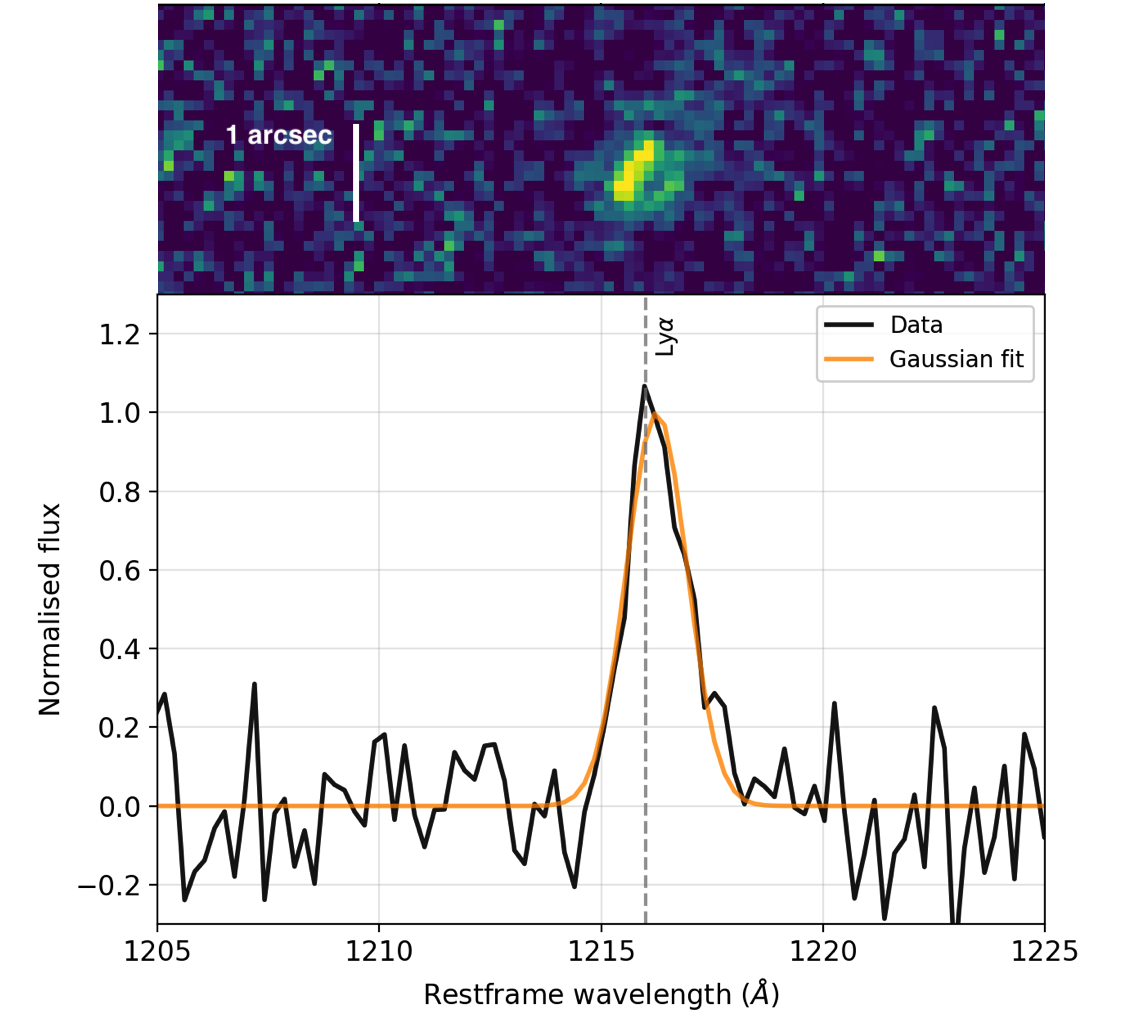}
	\vspace{-10pt}
	\caption{Rest-frame 1D spectrum showing the asymmetric Ly$\alpha$ line profile at a redshift of $z=5.720$. Also shown is the best-fit Gaussian to the emission line. The peak of the fitted Gaussian is slightly redder than the peak of the line, suggesting asymmetry in the emission line. This is also clear from the excess towards the redder parts of the Gaussian. \textit{Top:} The 2D GMOS spectrum showing the detected Ly$\alpha$ line. The spatial extent of the emission is roughly 1 arcsecond, which is also the aperture size used to extract the 1D spectrum.}
	\label{fig:rest_spec}
\end{figure}

We identify the single emission line detected in the GMOS spectrum as Ly$\alpha$ $\lambda 1216$, giving a redshift of $z=5.720\pm0.001$, which is shown in Figure \ref{fig:rest_spec}. Other plausible identifications of this emission line could be [O \textsc{iii}] $\mathrm{\lambda} 5007$, giving a redshift of $z\approx0.63$ or H$\alpha$ $\mathrm{\lambda} 6563$ at $z\approx0.25$. These can be ruled out given the non-detection of other bright lines expected in the wavelength range covered. For example, if the observed line is [O \textsc{iii}] at $z=0.63$, then the [O \textsc{ii}] $\lambda 3727$ line would be observed at $\sim6075$ \AA$ $, unless the galaxy has a particularly high [O \textsc{iii}]/[O \textsc{ii}] ratio as in the case of very low metallicity objects. An unresolved [O \textsc{ii}] $\lambda \lambda 3726, 3729$ doublet at a redshift of $z\approx1.2$ could be a possibility, but the absence of other expected UV/optical lines common in AGN and radio galaxy spectra, such as C \textsc{ii}] $\lambda 2326$ or Mg \textsc{ii} $\mathrm{\lambda \lambda} 2797,2803$, which are on average a factor of $2-4$ times fainter than [O \textsc{ii}] \citep{deb01}, makes this possibility unlikely. Another possibility could be C \textsc{iv} $\lambda 1549$ giving $z=4.27$, but this can be ruled out because in this case Ly$\alpha$, which is often much brighter (for example, there are no radio galaxies in \citet{deb01} with Ly$\alpha$ flux lower than C \textsc{iv} $\lambda 1549$ flux), would be expected at $\sim6410$ \AA$ $ and we do not see any signs of an emission line in that region, which is free from bright sky lines.

We fit a Gaussian to the emission line (shown in Figure \ref{fig:rest_spec}) to measure an integrated line flux of $F_{Ly\alpha} = 1.6 \pm 0.2\times10^{-17}$ erg s$^{-1}$ cm$^{-2}$. The total measured Ly$\alpha$ luminosity is $L_{{Ly\alpha}} = 5.7 \pm 0.7 \times 10^{42}$ erg s$^{-1}$. The full width at half maximum (FWHM) after correcting for the instrumental FWHM is $370 \pm 30$ km s$^{-1}$. Since no continuum is detected in the spectrum (down to $1\sigma$ depth of $4.0 \times 10^{-19}$ erg s$^{-1}$ cm$^{-2}$), we can only put a lower limit on the rest-frame equivalent width (EW) of the line, EW$_{0}$ $> 40$ \AA $ $.  Table \ref{tab:line} presents a summary of emission line measurements for TGSS J1530+1049.

\subsection{Skewness and equivalent width}
To further confirm our redshift determination, we quantify the asymmetry of the emission line following the prescriptions laid out by \citet{kas06}, by calculating the S-statistic and the weighted skewness parameter. A measure of the skewness of the emission line is particularly useful when dealing with spectra with a single emission line and can help differentiate Ly$\alpha$ emission at high redshifts from [O \textsc{ii}], [O \textsc{iii}] or H$\alpha$ emission from lower redshift galaxies \citep{rho03, kur04, kas06}. We measure the skewness $S = 0.31 \pm 0.14$ and the weighted skewness $S_w = 6.44 \pm 2.97$ \AA $ $, which are consistent with what is observed for confirmed Ly$\alpha$ emitters at high redshift \citep{kas06, kas11,mat17}.

To check what possible values of skewness could be obtained from an unresolved [O \textsc{ii}] doublet, we simulated the doublet with all possible ratios ($0.35<j_{\lambda3729}/j_{\lambda3726}<1.5$), convolved with the instrument resolution. We find that the skewness measured for the emission line seen in the spectrum ($S=0.31$) is only possible for $j_{\lambda3729}/j_{\lambda3726} < 0.7$. These line ratios correspond to the high electron density regime when the line would be collisionally de-excited, and hence unlikely to be as strong as observed, with previous studies of the [O \textsc{ii}] doublet in high-z galaxies \citep{ste14,shi15,san16} also finding much higher line ratios on average. This helps drive the interpretation of the observed emission line more towards a Ly$\alpha$ at high redshift. 

Further, an EW $> 40$ \AA $ $ for an [O \textsc{ii}] line originating from a presumably massive radio galaxy at $z\approx1.2$ would be at the extreme end of the EW distribution \citep{bri15}, including for radio-loud quasars \citep{kal12}. This EW value is also incompatible with the line ratios that would give rise to the observed skewness, as in regions of very high electron densities the [O {\textsc{ii}}] line is expected to be weaker due to collisional de-excitation. Therefore, we can practically rule out the [O \textsc{ii}] doublet as a possible identification of this emission line. An EW$_0$ $>40$ \AA $ $, however, is typical for Ly$\alpha$ emission seen in galaxies at $z\approx5.7$ \citep[see][for example]{kas11} and generally consistent with the $z\sim6$ galaxy population \citep{deb17}. 

\subsection{K-z relation for radio galaxies}
Finally, a strong indicator of a high-redshift nature of the host galaxy is the non-detection in $K$ band down to a $3\sigma$ limiting magnitude of 22.4 (Figure \ref{fig:kband}) using aperture photometry at the peak pixel of the radio emission. For comparison, TN J0924$-$2201 at $z=5.2$ has a magnitude of $K=21.3 \pm 0.3$ and our measurement of $K>22.4$ is consistent with $z>5$ and helps rule out lower redshifts owing to the $K-z$ relation for radio galaxies (note that the luminosity and the spectral index rule out that it is a star-forming galaxy). We expand upon this point in Section \ref{subsec:kband}. The additional non-detection in $J$ band down to a $3\sigma$ limit of 24.3 further favours a high redshift galaxy and supports the argument that the line we see is indeed Ly$\alpha$ and not [O \textsc{ii}].

\section{Discussion}
\label{sec:discussion}
\subsection{Emission line measurements}
\label{subsec:line}
\begin{table}
\centering
\caption{Spectroscopic redshift and emission line measurements for TGSS J1530+1049 through GMOS spectroscopy.}
	\begin{tabular}{l r}
    \hline
    Property	&	Measurement	\\
    \hline \hline
    $z_{\textrm{spec}}$	&	$5.720\pm0.001$	\\
    $F_{\textrm{Ly}\alpha}$	&	$1.6 \pm 0.2 \times 10^{-17}$ erg s$^{-1}$ cm$^{-2}$	\\
    $L_{\textrm{Ly}\alpha}$	& $5.7 \pm 0.7 \times 10^{42}$ erg s$^{-1}$ \\
    FWHM	&	$370 \pm 30$ km s$^{-1}$	\\
    EW$_{\textrm{obs}}$	&	$>40$ \AA	\\
    \hline
    \end{tabular}
\label{tab:line}
\end{table}

The Ly$\alpha$ luminosity and FWHM measured for TGSS J1530+1049 are lower than what is seen for typical HzRGs at $z>4$ \citep[see][for examples]{spi95, deb99, vbr99, mil08} and more consistent with those measured for `non-radio' Ly$\alpha$ emitting galaxies (LAEs) at this redshift \citep{rho03, ouc08, kas11, lid12, mat17}. However, the FWHM for TGSS J1530+1049 is consistent with that of a very faint radio galaxy VLA J123642$+$621331, with a 1.4 GHz flux density of $S_{\textrm{1.4 GHz}} = 0.47$ mJy, discovered at $z=4.424$ \citep{wad99}. This galaxy has a FWHM of $\approx420$ km s$^{-1}$ and a Ly$\alpha$ luminosity $\approx 2 \times 10^{42}$ erg s$^{-1}$, which is weaker than TGSS J1530+1049. VLA J123642$+$621331 is however, not detected in TGSS at 150 MHz down to a noise level of 3.5 mJy beam$^{-1}$, suggesting a relatively flat spectral index or a spectral turnover at low radio frequencies. We present some comparisons of the Ly$\alpha$ properties we measure for TGSS J1530+1049 with other HzRGs at $z>4$ and also non-radio LAEs at $z=5.7$ in Table \ref{tab:comparison}.
\begin{table*}
\centering
\caption{Comparison of Ly$\alpha$ emission line properties of TGSS J1530+1049 reported in this paper with some of the known radio galaxies at $z>4$, and the much fainter radio galaxy at $z=4.42$ (all marked as RG), in addition to several confirmed LAEs at $z\approx5.7$ from the literature.}
	\begin{tabular}{l c c c c r}
	\hline
					&			&	$F_{Ly\alpha}$						&	$L_{Ly\alpha}$ 				&	FWHM$_{Ly\alpha}$		&	 \\
	Name			&	$z$		& ($\times 10^{-17}$ erg s$^{-1}$ cm$^{-2}$)	& ($\times 10^{42}$ erg s$^{-1}$)	&	(km s$^{-1}$)			&	Reference \\
	\hline \hline
	TGSS J1530+1049			&	5.72	&	1.6	&	5.7	&	370		&	This work		\\
	TN J0924$-$2201 (RG)	&	5.19	&	3.4	& 9.6	&	1500	&	\citep{vbr99}	\\
    J163912.11$+$405236.5 (RG) & 4.88	& 18.5	&	47.0	&	1040	&	\citep{jar09} \\
    VLA J123642$+$621331 (RG) &	4.42 	&	0.6	&	2.0		&	420		&	\citep{wad99}	\\
    \hline
	LALA J142546.76$+$352036.3 &	5.75	&	1.6	&	6.7	&	360		&	\citep{rho03}	\\
	S11 5236				&	5.72	&	14.9	&	9.1	&	200		&	\citep{lid12}	\\
    SDF J132344.8$+$272427	&	5.72	&	1.1		&	3.7	&	366		&	\citep{kas11}	\\
    HSC J232558+002557		&	5.70	&	3.6		&	12.6	&	373	&	\citep{shi18}	\\
    SR6						 &	5.67 	&  	7.6	&	25.0	&	236		&	\citep{mat17}	\\
	\hline		
	\end{tabular}
\label{tab:comparison}
\end{table*}

A statistical sample of radio galaxies at $z\sim6$ is needed to understand whether they are more like LAEs at high redshift or whether a majority of them continue being very different systems, surrounded by extremely overdense regions and forming stars intensively. The relatively underluminous Ly$\alpha$ would be one signature of a significantly neutral intergalactic medium (IGM) during the late stages of the EoR. Weaker Ly$\alpha$ emission may also be caused by significant absorption in a cold and dusty medium surrounding the radio galaxy. The presence of cold gas and dust has been reported in many HzRGs, including TN J0924$-$2201 \citep[see][for example]{kla05} and dedicated observations to look for molecular gas and dust in a statistically significant sample of radio galaxies at $z>5$ are required to better characterise their surrounding medium.
	
\subsection{Radio properties}
TGSS J1530+1049 has a flux density of 170 mJy at a frequency of 150 MHz and 7.5 mJy at 1.4 GHz \citep{sax18}. Using the standard $K$-corrections in radio astronomy and assuming a constant spectral index of $\alpha=-1.4$, we calculate a rest-frame radio luminosity of $\log L_{\textrm{150 MHz}} = 29.1$ and $\log L_{\textrm{1.4 GHz}} = 28.2$ W Hz$^{-1}$, which places this source at the most luminous end of the radio luminosity function at this epoch \citep{sax17}. For comparison, TN J0924$-$2201 has a $K$-corrected radio luminosity of $\log L_{\textrm{1.4 GHz}} = 29.3$ W Hz$^{-1}$ using a spectral index of $\alpha=-1.6$ \citep{vbr99}. TGSS J1530+1049 is close to an order of magnitude fainter than TN J09224$-$2201 at 1.4 GHz, but remains by far the brightest radio source observed this close to the end of the epoch of reionisation. 

The deconvolved angular size determined by \citet{sax18} at 1.4 GHz for TGSS J1530+1049, which remains unresolved, is 0.6 arcseconds, which translates to a linear size of 3.5 kpc. This size is smaller than the size of TN J0924$-$2201 \citep{vbr99} and in line with predictions at $z\sim6$ from \citet{sax17}, as radio galaxies in the early Universe are expected to be young and very compact \citep{blu99}. In Table \ref{tab:radiocomp} we compare the radio properties of TGSS J1530+1049 with all currently known radio galaxies at $z>4$. This was done by querying the TGSS ADR1 catalog to determine flux densities for all $z>4$ radio galaxies at 150 MHz, which were then used to calculate radio powers using the standard $K$-corrections. We find that TGSS J1530+1049 is comparable to many of the $z>4$ radio galaxies when looking at radio properties alone.
\begin{table*}
\centering
\caption{A comparison of the radio properties of TGSS J1530+1049 with other known radio galaxies at $z>4$. Flux densities at 150 MHz are measured from the TGSS catalog.}
	\begin{tabular}{l c c c c c c r}
    \hline
    			&		&	$S_{150}$ 	&	$\log L_{\textrm{150}}$ &	$S_{1400}$	&							&	Size 	&			\\
    Name		&	$z$	&	(mJy)		&	(W Hz$^{-1}$)			&	(mJy)		&	$\alpha^{150}_{1400}$	&	(kpc)	&	Reference	\\	
    \hline \hline
    TGSS J1530+1049		& 5.72	&	170		&	29.1	&	7.5	&	$-1.4$	&	3.5		&	This work \\
    \hline
    TN J0924$-$2201	&	5.19	&	760		&	29.6	&	71.1	&	$-1.1$	&	7.4		&	\citep{vbr99} \\
    J163912.11$+$405236.5 & 4.88 &	103		& 	28.2	&	22.5	&	$-0.7$	&	$-$		& \citep{jar09} \\
    RC J0311$+$0507	&	4.51	&	5981	&	30.3	&	500.0	&	$-1.1$	&	21.1	&	\citep{par14} \\
    VLA J123642$+$621331 & 4.42 & 	undetected 	&	$-$	&	0.5	&	$-$		&	$-$		&	\citep{wad99} \\
    6C 0140$+$326	&	4.41	&	860		&	29.4	&	91.0	&	$-1.0$	&	17.3	&	\citep{raw96} \\
    8C 1435$+$635	&	4.25	&	8070	&	30.4	&	497.0	&	$-1.2$	&	21.1	&	\citep{lac94} \\
    TN J1123$-$2154	&	4.11	&	512		&	29.1	&	49.3	&	$-1.0$	&	5.5		&	\citep{reu04} \\
    TN J1338$-$1942	&	4.10	&	1213	&	29.5	&	120.8	&	$-1.0$	&	37.8	&	\citep{reu04} \\
    \hline
    \end{tabular}
\label{tab:radiocomp}
\end{table*}

TGSS J1530+1049 has a spectral index of $\alpha^{\textrm{\tiny{150 MHz}}}_{\textrm{\tiny{1.4 GHz}}} = -1.4$, which is ultra-steep but flatter than TN J0924$-$2201 at $z=5.19$, which was selected because of its spectral index of $\alpha^{\textrm{\tiny{365 MHz}}}_{\textrm{\tiny{1.4 GHz}}} = -1.6$. Interestingly, at lower radio frequencies the spectral index of TN J0924$-$2201 appears to flatten dramatically. The 150 MHz flux density measured in TGSS \citep{int17} for TN J0924$-$2201 is $760 \pm 76$ mJy, giving a low frequency spectral index $\alpha^{\textrm{\tiny{150 MHz}}}_{\textrm{\tiny{365 MHz}}} = -0.16$. If the spectral index were to be calculated only using the flux densities at frequencies of 150 MHz and 1.4 GHz, the inferred spectral index would be $\alpha^{\textrm{\tiny{150 MHz}}}_{\textrm{\tiny{1.4 GHz}}} = -1.06$, making it not strictly ultra-steep ($\alpha < -1.3$). This implies that in a search for ultra-steep spectrum radio sources using data at 150 MHz and 1.4 GHz, such as \citet{sax18}, TN J0924$-$2201 would be missed entirely. Indeed, an indication of a generally flatter spectral index at lower frequencies is visible in the spectral indices calculated for known $z>4$ radio galaxies between frequencies of 150 MHz and 1.4 GHz in Table \ref{tab:radiocomp}. A large majority of these radio sources were selected for having an ultra-steep spectral index in a higher frequency range, but appear to have a flatter spectral index when calculated between 150 MHz and 1.4 GHz.

Spectral flattening or even a turnover at low radio frequencies is expected in radio galaxies at increasingly higher redshifts due to: a) Inverse Compton (IC) losses due to the denser cosmic microwave background that affect the higher frequencies and result in a steeper high frequency spectral index, and b) free-free or synchrotron self absorption due to the compact sizes of radio sources at high redshifts that can lead to a turnover in the low frequency spectrum \citep[see][and references therein]{cal17}. \citet{sax18} have reported evidence of flattening of the low-frequency spectral index in candidate HzRGs and observations at intermediate radio wavelengths for sources like TGSS J1530+1049 are essential to measure spectral flattening and constrain various energy loss mechanisms that dominate the environments of radio galaxies in the early Universe. Additionally, search techniques for radio galaxies at even higher redshifts could be refined by possibly using radio colours instead of a simple ultra-steep spectral index selection. The LOFAR Two Metre Sky Survey (\citealt{shi17}, Shimwell et al. submitted) will eventually provide in-band spectral indices at 150 MHz and could potentially be used to identify HzRG candidates more efficiently.

We also draw attention towards the radio galaxy J163912.11$+$405236.5 at $z=4.88$ \citep{jar09}, that has a spectral index of $\alpha^{\textrm{\tiny{325 MHz}}}_{\textrm{\tiny{1.4 GHz}}} = -0.75$ and is not an ultra-steep spectrum radio source. Interestingly, there is evidence of spectral flattening at lower frequencies with a 150 MHz flux density of 103.5 mJy, giving a spectral index $\alpha^{\textrm{\tiny{150 MHz}}}_{\textrm{\tiny{325 MHz}}} = -0.56$, which is flatter than that at higher frequencies. This source was targeted for spectroscopic follow-up owing to the faintness of its host galaxy at $3.6$ $\mu m$. The very faint radio galaxy VLA J123642$+$621331 at $z = 4.42$ \citep{wad99} is also not strictly an ultra-steep spectrum source at high radio frequencies ($\alpha^{\textrm{\tiny{1.4 GHz}}}_{\textrm{\tiny{4.8 GHz}}} = -0.94$) and is too faint to be detected in TGSS. This source was also selected based on its optical and infrared faintness, suggesting that a considerable fraction of HzRGs may not be ultra-steep at all and therefore, be missed in samples constructed using the ultra-steep spectrum selection technique. 

Indeed \citet{ker12} have shown that selecting infrared-faint radio sources (IFRS) could be more efficient at isolating HzRGs from large samples when compared to radio selection alone. This has been confirmed observationally through many previous studies \citep{nor06, col14, her14, mai16, sin17}. However, the caveat is that deep infrared photometry over large sky areas is required to effectively implement such a selection, which can be expensive. The recently concluded UKIRT Hemisphere Survey (UHS; \citealt{dye18}) has the potential to be extremely useful in the identification of promising HzRG candidates in the Northern Hemisphere, particularly from the LOFAR surveys (\citealt{shi17}, Shimwell et al. submitted), as a combination of a relaxed spectral index steepness criterion and infrared-faintness could be more effective at isolating HzRGs.

\subsection{Stellar mass limits}
\label{subsec:kband}
The non-detection of the host galaxy down to a $3\sigma$ depth of $K=22.4$ can be used to set limits on the stellar mass for TGSS J1530+1049 using simple stellar population synthesis modelling. To do this, we make use of the \textsc{python} package \textsc{smpy}\footnote{\url{https://github.com/dunkenj/smpy}}, which is designed for building composite stellar populations in an easy and flexible manner, allowing for synthetic photometry to be produced for single or large suites of models \citep[see][]{dun15}. To build stellar populations, we use the \citet{bc03} model with a \citet{cha03} initial mass function (IMF) and solar metallicity \citep{wil03}, a formation redshift $z_f = 25$ and assume a maximally old stellar population that has been forming stars at a constant rate \citep{lac00}. We follow the \citet{cal00} law for dust attenuation and use values of $A_v = 0.15$ (moderate extinction) and $0.5$ mag (dusty), which are commonly seen in massive galaxies at $5<z<6$ \citep{mcl06}. The synthetic photometry is produced for different stellar masses, which we then convolve with the $K$ band filter to calculate apparent $K$ magnitudes over a redshift range $0-7$.
\begin{figure}
\centering
	\vspace{-10pt}
	\includegraphics[scale=0.45]{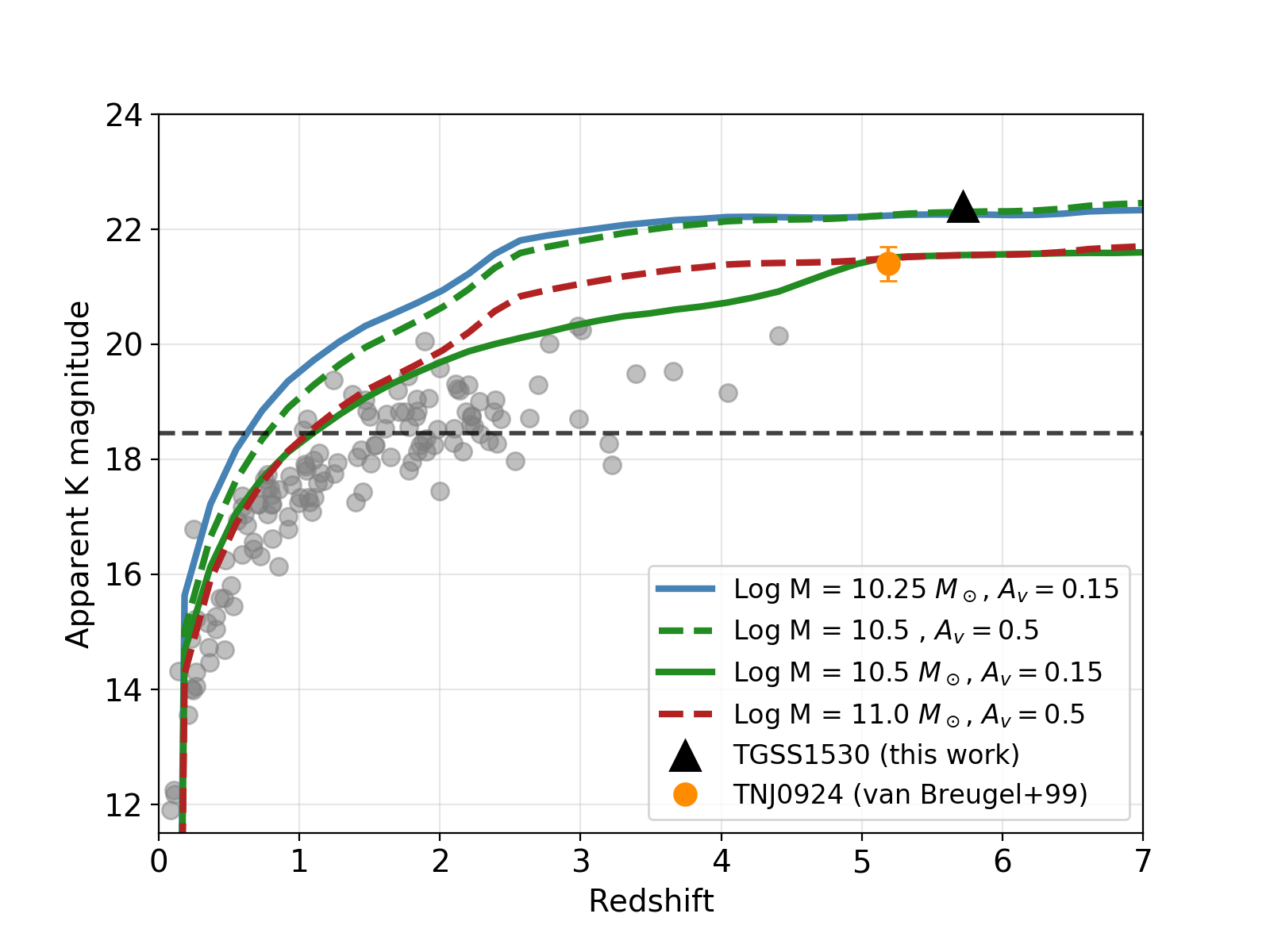}
    \vspace{-10pt}
	\caption{The `$K-z$' diagram for radio galaxies, showing stellar mass limits derived from stellar population synthesis modelling for TGSS J1530+1049 (black triangle). The $K$-band $3\sigma$ limit gives $M_{\textrm{stars}} < 10^{10.25}$ $M_\odot$ for $A_v = 0.15$ mag, and $M_{\textrm{stars}} < 10^{10.5}$ for $A_v = 0.5$ mag. Also shown are $K$-band magnitudes and redshifts for known radio galaxies in the literature (grey points; see text), with TN J0924$-$2201 at $z=5.19$ (orange circle). The $K$-band limits for TGSS J1530+1049 further help exclude lower redshift measurements from incorrect line identification.}
	\label{fig:kz}
\end{figure}

The $K$ magnitude limit for TGSS J1530+1049 fits well with a stellar mass limit of $M_{\textrm{stars}} < \sim 10^{10.25}$ $M_\odot$ for $A_v = 0.15$ mag, and $M_{\textrm{stars}} < \sim 10^{10.5}$ for $A_v = 0.5$ mag. We note here that thanks to the excellent seeing for $K$-band observations ($0.6-0.8$ arcseconds), and since at $z\sim 6$ the host galaxy is expected to be small, any aperture correction is only expected to be at the level of a few tenths of a magnitude at most, or $0.1-0.2$ in the logarithmic stellar mass, which is smaller than the uncertainty from dust extinction corrections. 

We find that the stellar mass limits we infer are in agreement with the $J$ band $3\sigma$ limit from LBT. The photometry predicted by the models in the optical bands from PS1 ($g$, $r$, $i$, $z$, $y$) is also consistent with the non-detections that we report. This stellar mass limit places TGSS J1530+1049 towards the $>M^*$ end of the galaxy stellar mass function at $z\sim6$ \citep[see][for example]{dun14}. For comparison, we show the apparent $K$ band magnitudes of other radio galaxies in the literature, taken from the 3CRR, 6CE, 6C* and 7C$-$I/II/III samples \citep{wil03}, in Figure \ref{fig:kz}. Also shown is the $K$ magnitude for TN J09224$-$2201 at $z=5.2$ \citep{vbr99}, which is best fit with a stellar mass of $10^{10.5}$ $M_\odot$ for $A_v = 0.15$ mag and $10^{11}$ $M_\odot$ for $A_v = 0.5$ mag. 

We also show the $K$-band magnitude limit for the UKIDSS LAS as a dashed black line in Figure \ref{fig:kz}. TGSS J1530+1049 was initially selected due to its non-detection in LAS. However, these magnitude limits alone were not sufficient to constrain the very high redshift nature of the host galaxy. With deeper LBT observations in $K$ band, we show that TGSS J1530+1049 follows the trend in the $K-z$ plot for radio galaxies. It is also clear that a low redshift solution that would arise if the detected emission line in Section \ref{sec:redshift} is not Lyman alpha (for example, $z\approx1.2$ if the line is [O \textsc{ii}]) would be hard to explain using galaxy evolution models with the inputs and assumptions outlined above and those generally used to model radio galaxy spectra \citep{ove09}.

A stellar mass limit of $M_{\star} < 10^{10.5}$ $M_\odot$ for TGSS J1530+1049 (based on the outlined assumptions) is almost an order of magnitude lower than the sample of radio galaxies studied by \citet{roc04}, where the highest stellar masses are seen for radio galaxies in the redshift range $2-4$. This suggests that TGSS J1530+1049 is still evolving and is in the process of building up its stellar mass. This interpretation is in line with TGSS J1530+1049 being a relatively young radio galaxy owing to its small radio size, and lower stellar masses may be expected from radio galaxies close to or into the epoch of reionisation. Robust detections at optical and infrared wavelengths are required to properly characterise the galaxy spectral energy distribution in order to understand better the star formation history of TGSS J1530+1049.

\section{Conclusions}
\label{sec:conclusion}
In this paper we have presented the discovery of the highest redshift radio galaxy, TGSS J1530+1049, at $z = 5.72$. The galaxy was initially selected at 150 MHz from TGSS \citep{sax18} and was assigned a high priority for spectroscopic follow-up owing to its compact morphology and faintness at optical and near-infrared wavelengths. The conclusions of this study are listed below:

\begin{enumerate}
	
    \item Long-slit spectroscopy centered at the radio position of the source revealed an emission line at 8170 \AA$ $, which we identify as Lyman alpha at $z = 5.720$. We rule out alternative line IDs owing to the absence of other optical/UV lines in the spectrum, the asymmetrical nature of the emission line characteristic of Lyman alpha at high redshifts that we quantify using the skewness parameter and the high observed equivalent width of the emission line.
     \item Deep $J$ and $K$ band imaging using the Large Binocular Telescope led to no significant detection of the host galaxy down to $3\sigma$ limits of $K>22.4$ and $J>24.4$. The limits in $K$ can be used as an additional constraint on the redshift, owing to the relation that exists between $K$ band magnitude and redshift of radio galaxies. The magnitude limit is consistent with $z>5$, practically ruling out a redshift of $z\approx1.2$ that would be expected if the emission line were an unresolved [O \textsc{ii}] $\lambda \lambda 3726, 3729$ doublet, which is the most likely alternative line identification.
    \item The emission line is best fitted with a skewed Gaussian, giving an integrated line flux of $F_{Ly\alpha} = 1.6\times10^{-17}$ erg s$^{-1}$ cm$^{-2}$, a Ly$\alpha$ luminosity of $5.7 \times 10^{42}$ erg s$^{-1}$, an equivalent width of EW $>40$ \AA$ $ and a FWHM of $370$ km s$^{-1}$. These values are more consistent with those observed in non-radio Lyman alpha emitting galaxies at this redshift and much lower than those corresponding to typical radio galaxies at $z>4$.
    \item The radio luminosity calculated at 150 MHz is $\log L_{\textrm{150}} = 29.1$ W Hz$^{-1}$, which places it at the most luminous end of the radio luminosity function at this epoch. The deconvolved angular size is 3.5 kpc, which is in line with the compact morphologies expected at high redshifts. We find that the radio properties of TGSS J1530+1049 are comparable to other known radio galaxies at $z>4$ but the compact size suggests that it is a radio galaxy in an early phase of its evolution. A joint study of the Ly$\alpha$ halo and the radio size of this source may provide one of the earliest constraints on the effects of radio-mode feedback.
    \item We use the $K$ band limit to put constraints on the stellar mass estimate using simple stellar population synthesis models. Assuming a constant star formation history and a maximally old stellar population, we derive a stellar mass limit of $M_{\textrm{stars}} < \sim 10^{10.25}$ $M_\odot$ for $A_v = 0.15$ mag, and $M_{\textrm{stars}} < \sim 10^{10.5}$ for $A_v = 0.5$ mag. These limits are almost an order of magnitude lower than typical radio galaxy masses in the redshift range $2-3$ and suggest that TGSS J1530+1049 may still be in the process of assembling its stellar mass, which is in line with it being a relatively young radio galaxy.
\end{enumerate}
    
An effective application of deep radio surveys covering very large areas on the sky has been demonstrated by this discovery of the first radio galaxy at a record distance after almost 20 years. With the more sensitive, large area surveys currently underway with LOFAR (LoTSS; \citealt{shi17}, Shimwell et al. submitted), there is potential to push searches for radio galaxies to even higher redshifts. Discovery of even a single bright radio galaxy at $z>6$ would open up new ways to study the epoch of reionisation in unparalleled detail, through searches for the 21cm absorption features left behind by the neutral hydrogen that pervaded the Universe at high redshifts.

\section*{Acknowledgements}
The authors thank the referee for useful comments and suggestions. AS would like to thank Jorryt Matthee, David Sobral, Reinout van Weeren and Nobunari Kashikawa for fruitful discussions. AS, HJR and KJD gratefully acknowledge support from the European Research Council under the European Unions Seventh Framework Programme (FP/2007-2013)/ERC Advanced Grant NEWCLUSTERS-321271. RAO and MM received support from CNPq (400738/2014-7, 309456/2016-9) and FAPERJ (202.876/2015). PNB is grateful for support from STFC via grant ST/M001229/1. IP acknowledges funding from the INAF PRIN-SKA 2017 project 1.05.01.88.04 (FORECaST).

This paper is based on results from observations obtained at the Gemini Observatory, which is operated by the Association of Universities for Research in Astronomy, Inc., under a cooperative agreement with the NSF on behalf of the Gemini partnership: the National Science Foundation (United States), the National Research Council (Canada), CONICYT (Chile), Ministerio de Ciencia, Tecnolog\'{i}a e Innovaci\'{o}n Productiva (Argentina), and Minist\'{e}rio da Ci\^{e}ncia, Tecnologia, Inova\c{c}\~{a}o e Comunica\c{c}\~{o}es (Brazil). This paper also contains data from the Large Binocular Telescope (LBT), an international collaboration among institutions in the United States, Italy and Germany. LBT Corporation partners are: The University of Arizona on behalf of the Arizona university system; Istituto Nazionale di Astrofisica, Italy; LBT Beteiligungsgesellschaft, Germany, representing the Max-Planck Society, the Astrophysical Institute Potsdam, and Heidelberg University; The Ohio State University, and The Research Corporation, on behalf of The University of Notre Dame, University of Minnesota, and University of Virginia.

This work has made extensive use of \textsc{ipython} \citep{per07}, \textsc{astropy} \citep{ast13}, \textsc{aplpy} \citep{apl}, \textsc{matplotlib} \citep{plt} and \textsc{topcat} \citep{top05}. This work would not have been possible without the countless hours put in by members of the open-source developing community all around the world. 



\bibliographystyle{mnras}
\bibliography{hzrg_paper} 




\bsp	
\label{lastpage}
\end{document}